# Strong Disorder and High Temperature Superconductivity

J. C. Phillips

Dept. of Physics and Astronomy, Rutgers University, Piscataway, N. J., 08854-8019

**Abstract**

ARPES and wide area, high resolution STM studies of a micaceous cuprate high temperature superconductor have shown strong nanoscale gap disorder, but the question of whether this disorder is intrinsic and necessary for HTSC, or merely incidental to either this material, or even only its surface, appears to be an open one. We present the case that it is merely incidental, and then review the evidence that strong disorder is actually an essential factor in generating this unprecedented phenomenon.

## 1. Introduction

The discovery of HTSC in cuprates 20 years ago astonished the world of condensed matter physics. The search for higher $T_c$ 's had virtually ended in the 1970's, and there were strong phenomenological arguments that $T_c$ would always be limited to ~ 30K because of lattice instabilities [1]. Now theory had to regroup and reconsider the problem afresh. Two possibilities existed: one could (A) try to extend the earlier BCS theory by considering the special properties of cuprates, or one could (B) abandon BCS and electron-phonon interactions altogether in favor of some other kind of interaction. I have argued that (A) is possible, and is consistent with the basic rules of logic (Ockham's razor), but I have been very much in the minority.

Recently a series of powerful experiments, based mainly on ARPES and wide area, high resolution STM, have shown that the cuprates have complex atomic and electronic structures, so that the task of theory appears to be one of analyzing how these complex structures promote HTSC. Some people have criticized this view as follows:

"Phillips claims that $T_c$ is high because cuprate stuffs are very messy crystals. This sounds to me like an opinion of a person who has not attended a high $T_c$ meeting during the last ten years. Surely, in the late 1980's this was a respectable idea. In the beginning the crystal growers had barely any control over the quality of the material and every high Tc crystal was a mess. However, the crystal growers have worked hard, very hard, over the last twenty years and in the mean time it is crystal clear that some cuprate crystals can be very clean (123), while others are to a greater or lesser degree messy (2212, etc), while some are intrinsically even very messy (214). The point is that there is no clear-cut correlation between crystalline dirt and superconductivity. To the contrary, a strong case is developing that cleanliness is very good for the superconductivity. At the least, I've been impressed by detailed studies by the materials people showing that Tc's in 214 are exceptionally low because of its intrinsic messy crystal. People are even giving names to the particular types of dirt being bad for superconductivity."

On the face of it, this sounds like the opinion of an expert. In my view, it is naïve and simplistic. However, it is not new or original: this represents an opinion that is shared by many "experts" (especially theorists) in the field, and I have often heard it before. Scientists who have not followed the field in detail find this argument quite persuasive. A striking feature of the argument is that it does not address the complex issues raised by the ARPES and STM data. In fact, one gets the very strong impression that these issues – the subjects of many papers by me – have been completely dismissed by such "experts" as artifacts of poorly prepared samples. This dismissive approach is certainly the easiest way of confronting complexity, but it is contradicted by the internal consistency of the experiments themselves, especially the phase diagrams that have been obtained.

**2. Unconventional Trends Reflect Complexity, Not Exotic Interactions**

The details of the experiments are presented and analyzed phenomenologically in the their original papers, which I have cited extensively; my own views (the self-organized, off-lattice (dopant-centered) network model, type (A)) can be found by searching SCI and

cond-mat/. The phase diagrams exhibit many anomalies. In general the high values of $T_c$ might be explained by (B) exotic interactions, but it is striking that after 20 years no one has been able to find a satisfactory model, though many have tried [2,3]. (The problem of finding an exotic lattice replacement for BCS is so hard that, even if possible, it might take 100 years or even longer, but the failures to date have certainly left the proponents of this approach with very little to stand on.) However - and this is very important – the complexities of the problem are not limited merely to anomalously high $T_c$ 's. The most important anomalies mainly show up as extremely peculiar chemical trends in the phase diagrams, and it is here that the era of high-quality samples has had its greatest impact – not, as suggested in the quote above, in the study of truly incidental and irrelevant defects, or variations in $T_c$ 's between different layered structures – these are interesting in themselves, and worthwhile, but they do not go to the heart of the matter, which is the phase diagram complexities themselves.

Let us review these major anomalies:
(1) The canonical phase diagram reveals an intermediate phase sandwiched between an insulator and a Fermi liquid. Only the intermediate phase is superconductive, and its anomalous properties (such as a nearly linear planar resistivity $\rho(T)$) persist to temperatures far above $T_c$. Moreover, the maximum in $T_c$ occurs near the center of this phase, where $\rho(T)$ is almost exactly linear over a wide temperature range. Normally it would occur near the boundary with the insulating phase, where carrier densities are lowest, reducing screening of (A) the electron-phonon interaction, or (B) some kind of exotic interaction based on the Coulomb interaction. In other words, the nature of the canonical phase diagram cannot be explained by any simplistic homogeneous model, regardless of whether one uses (A) or (B). The complexity, and specifically the strong disorder, are essential to understanding the canonical phase diagram [4].
(2) Now we go on to the very sophisticated anomalies revealed by the detailed chemical trends found in ARPES and STM experiments. First, one has the ~ 3 nm larger and smaller (bimodal distribution) gap domains with a patchy spatial structure that nevertheless, after Fourier transformation, yields the same d wave gap anisotropy (albeit superimposed on a very large ("dark matter") background, which is still being analyzed) as is obtained directly in **k** space by ARPES. The bimodal distribution shows that the filling factors for the two gaps vary smoothly across the phase diagram. If this structure is merely a "messy" artifact, why do the two completely complementary experiments agree so well? Surface science is by now itself a sophisticated subject, and no surface scientist has come forward to challenge the intrinsic nature of these results.

(3) If the correlations with the overall canonical phase diagram were not enough, what about the ARPES Fermi arcs that evolve with doping? Perhaps these could be explained by some kind of order parameter theory, but step function changes in the relative intensities of the (10) and (11) gaps at optimal doping cannot be explained by any polynomial model – only an exponentially complex model can generate step functions. Similar step function anomalies are also observed in the relaxation of spectral holes at 1.5 eV, which is hardly something that can be explained using a continuum model.

(4) The diamagnetic anomalies that appear to be associated with the pseudogap have an onset temperature as large as $2T_c^{max}$. They cannot be explained by a continuum model.

(5) The isotope effect in $T_c$, whose absence (or at least reduction) near optimal doping was supposed to make (A) impossible, has turned out to be very strong in ARPES data, and its phase diagram is fully consistent with (A), after allowance is made for network self-organized complexity.

**3. The Bottom Line**

Most oxides are "messy" and complex, and it is difficult to separate these two aspects. However, thanks to 20 years of brilliant experimental work, we can now see that the cuprates are no longer "messy", but they are still complex, and they will always be so.